\documentclass[10pt]{article}
\usepackage{amsmath,amssymb,tikz}
\usepackage{graphicx}
 \allowdisplaybreaks
\def\be{\begin{equation}}
\def\ee{\end{equation}}
\def\ba#1\ea{\begin{align}#1\end{align}}
\def\bg#1\eg{\begin{gather}#1\end{gather}}
\def\bm#1\em{\begin{multline}#1\end{multline}}
\def\bmd#1\emd{\begin{multlined}#1\end{multlined}}

\def\e{\epsilon}

\def\fr{\frac}

\def\eq{\equiv}

\def\qqu{\qquad}

\def\({\left(}
\def\){\right)}
\def\[{\left[}
\def\]{\right]}

\def\zb{{\bar z}}

\long\def\symbolfootnote[#1]#2{\begingroup%
\def\thefootnote{\fnsymbol{footnote}}\footnote[#1]{#2}\endgroup}

\newcommand{\aei}{\it Max Planck Institute for Gravitational Physics
(Albert Einstein Institute)\\ Am M\"uhlenberg 1, 14476 Golm,
Germany}

\begin{document}
\thispagestyle{empty}
\begin{center}

~\vspace{20pt}

{\Large\bf A holographic proof of the universality of corner entanglement for CFTs}

\vspace{25pt}

Rong-Xin Miao\symbolfootnote[1]{Email:~\sf rong-xin.miao@aei.mpg.de}

\vspace{10pt}${}^\ast{}$\aei

\vspace{2cm}

\begin{abstract}
There appears a universal logarithmic term of entanglement entropy, i.e., $-a(\Omega) \log(H/\delta)$, for 3d CFTs when the entangling surface has a sharp corner. $a(\Omega)$ is a function of the corner opening angle and behaves as $a(\Omega\to \pi)\simeq \sigma (\pi-\Omega)^2$ and $a(\Omega\to 0)\simeq \kappa/\Omega$, respectively. Recently, it is conjectured that $\sigma/C_T=\pi^2/24 $, where $C_T$ is central charge in the stress tensor correlator, is universal for general CFTs in three dimensions. In this paper, by applying the general higher curvature gravity, we give a holographic proof of this conjecture. We also clarify some interesting problems. Firstly, we find that, in contrast to $\sigma/C_T$, $\kappa/C_T$ is not universal. Secondly, the lower bound $a_E(\Omega)/C_T$ associated to Einstein gravity can be violated by higher curvature gravity. Last but not least, we find that there are similar universal laws for CFTs in higher dimensions. We give some holographic tests of these new conjectures.  
\end{abstract}

\end{center}

\newpage
\setcounter{footnote}{0}
\setcounter{page}{1}

\tableofcontents

\section{Introduction}

The entanglement entropy (EE) of 3d CFTs takes the form
\begin{eqnarray}\label{3dEE}
S=B\ H/\delta -a(\Omega) \log(H/\delta)+ O(1)
\end{eqnarray}
where $\delta$ is a short-distance cutoff, $B$ is a constant and $H$ denotes the size of the entangling surface. The first term of eq.(\ref{3dEE}) is the 'area law' contribution to EE and the second logarithmic term appears only if the entangling surface has a sharp corner. For pure state, we have $a(\Omega)=a(2\pi-\Omega)$ due to the fact $S(V)=S(\bar{V})$. Besides, strong subadditivity and Lorentz invariance impose
\begin{eqnarray}\label{acondition}
a(\Omega)\ge 0, \ \ \partial_{\Omega}a(\Omega)\le 0, \ \  \partial^2_{\Omega}a(\Omega)\ge \frac{|\partial_{\Omega}a(\Omega)|}{\sin\Omega}\ \ \  \text{for}\ \ \Omega\le \pi.
\end{eqnarray}
 $a(\Omega)$ characterizes the CFTs and behaves as
\begin{eqnarray}\label{acondition1}
a(\Omega\to \pi)\simeq \sigma (\pi-\Omega)^2,\ \ \ a(\Omega\to 0)\simeq \kappa/\Omega
\end{eqnarray}
in the smooth and singular limits, respectively. 

Recently, it is conjectured that
\begin{eqnarray}\label{conjecture}
\sigma/C_T=\pi^2/24
\end{eqnarray}
 is a universal law for all CFTs in three dimensions \cite{Myers0}. Here $C_T$ is the central charge defined in the vacuum two-point function
\begin{eqnarray}\label{twopoint}
<T_{\mu\nu}(x)T_{\lambda\rho}(0)>=\frac{C_T}{|x|^{2d}}I_{\mu\nu,\lambda\rho}(x)
\end{eqnarray}
with $I_{\mu\nu,\lambda\rho}$ a dimensionless tensor fixed by symmetry. 

This conjecture has been tested in \cite{Myers0,Myers1} by studying some higher curvature holographic models, free scalars and fermions, and Wilson-Fisher fixed points of the $O(N)$ models with $N=1,2,3$ for $\frac{\pi}{2}$ corners. For recent developments, please refer to \cite{Elvang,Pang}. Let us briefly review the holographic models studied in \cite{Myers0,Myers1}. Consider the following action
\begin{eqnarray}\label{HDG0}
I&=&\frac{1}{16\pi G}\int d^4x \sqrt{g}\big{[} \frac{6}{L^2}+R+L^2(\lambda_1 R^2+\lambda_2 R_{\mu\nu}R^{\mu\nu}+\lambda_{\text{GB}}\mathcal{X}_4)\nonumber\\
 &&\ \ \ \ \ +L^4(\lambda_{3,0} R^3+\lambda_{1,1} R\mathcal{X}_4)+L^6(\lambda_{4,0} R^4+\lambda_{2,1}R^2\mathcal{X}_4 +\lambda_{0,2}\mathcal{X}^2_4)\big{]}
\end{eqnarray}
where $\mathcal{X}_4=R_{\mu\nu\rho\sigma}R^{\mu\nu\rho\sigma}-4R_{\mu\nu\rho}R^{\mu\nu}+R^2$ is the 4d Euler density. Since we are interested of the vacuum of the CFTs, we require the holographic model (\ref{HDG0}) has a pure $AdS_4$ solution. This imposes one constraint on the parameters $\lambda$. Ignoring the total derivatives and $K_aK^a$, \cite{Myers0,Myers1} find that the holographic entanglement entropy of the model (\ref{HDG0}) is equivalent to that of Einstein gravity 
\begin{eqnarray}\label{HEE0}
S&=&\alpha\int d^2y \frac{\sqrt{h}}{4 G}
\end{eqnarray}
up to a overall factor 
\begin{eqnarray}\label{alpha0}
\alpha=1-24\lambda_1-6\lambda_2+432\lambda_{3,0}+24\lambda_{1,1}-6912\lambda_{4,0}-576\lambda_{2,1}+O(\lambda^2).
\end{eqnarray}
As a result, we have $a(\Omega)=\alpha a_E(\Omega)$ where $a_E$ denotes the function for Einstein gravity.

Now let us discuss the central charge $C_T$ eq.(\ref{twopoint}). There is a standard holographic calculation of $C_T$ for Einstein gravity and one finds 
\begin{eqnarray}\label{CTE0}
C_{T,E}=\frac{3}{\pi^3}\frac{\tilde{L}^2}{G}
\end{eqnarray}
where $\tilde{L}$ is the radius of $AdS_4$. The situation is a little more complicated for higher curvature gravity. That is because, in addition to the usual massless spin-two gravition, massive modes and ghost modes with $M\sim 1/(\lambda L^2) $ also appear in higher curvature gravity. To suppress these modes, it is natural to work in the perturbative framework with $\lambda \ll 1$. Consider the metric fluctuations in the $AdS_4$ background together with the gauge $\bar{\nabla}^{\mu}h_{\mu\nu}=0$ and $g^{\mu\nu}h_{\mu\nu}=0$, we can derive the linearized Einstein equations as
\begin{eqnarray}\label{EOME}
-\frac{1}{2}[\bar{\Box}+\frac{2}{\bar{L}^2}]h_{\mu\nu}=8\pi G T_{\mu\nu}
\end{eqnarray}
Similarly, we can derive the linearized E.O.M for the holographic model  (\ref{HDG0})
\begin{eqnarray}\label{EOMHDG0}
-\frac{\alpha}{2}[\bar{\Box}+\frac{2}{\bar{L}^2}]h_{\mu\nu}-\frac{\lambda_2L^2}{2}[\bar{\Box}+\frac{2}{\bar{L}^2}]^2h_{\mu\nu}=8\pi G T_{\mu\nu}
\end{eqnarray}
Clearly, the second term of the above equation is suppressed near the physical pole, i.e. $[\bar{\Box}+\frac{2}{\bar{L}^2}]h_{\mu\nu}\sim 0$. Comparing eq.(\ref{EOMHDG0}) with eq.(\ref{EOME}), we notice that the effective Newton constant of the holographic model (\ref{HDG0}) is $G_{eff}=G/\alpha$. From eq.(\ref{CTE0}), we get $C_T=\alpha C_{T,E}$. Recall that we have $a(\Omega)=\alpha a_E(\Omega)$ from eq.(\ref{HEE0}). We finally obtain
\begin{eqnarray}\label{Myersresult}
\frac{a(\Omega)}{C_T}=\frac{a_E(\Omega)}{C_{T,E}}
\end{eqnarray}
which agrees with the conjecture (\ref{conjecture}).

Let us comment on the above holographic results of \cite{Myers0,Myers1}.

Firstly, the holographic model (\ref{HDG0}) is a combination of Einstein gravity, curvature-squared gravity and $f(Lovelock)$ gravity. Although it looks quite general, it is actually very special. As we know, in general, the extremal entropy surface of higher curvature gravity is no longer a minimal area surface. However, the bulk entangling surface of model (\ref{HDG0}) is still a minimal area surface. As a result, we have not only $\sigma/C_T$ (\ref{conjecture}) but also  $a(\Omega)/C_T$ (\ref{Myersresult}) universal. This is, however, not the case for free scalars and fermions. Thus, it is necessary to study more general higher curvature gravity models.

Secondly, \cite{Myers0,Myers1} have used the entropy formula proposed in \cite{Wall1} for $f(Lovelock)$ gravity. This entropy formula obeys the second law of thermodynamics for linearized perturbations of Killing horizons. However, it conflicts with the entropy formula proposed by \cite{Dong} at order $K^4$ ($K$ is the extrinsic curvature). Note that the entropy formula of \cite{Dong} also satisfies the linearized second law of theormodynamics \cite{Wall2}. Thus, it is necessary to check whether the holographic results of \cite{Myers0,Myers1} change if one use the entropy formula of \cite{Dong} instead of \cite{Wall1}.

We fill the above gaps in this paper. By studying the most general higher curvature gravity, we give a holographic proof of the conjecture (\ref{conjecture}) \cite{Myers0}. It seems impossible to find such a holographic proof because of the current limitations in understanding the holographic entanglement entropy for higher curvature gravity. Let us summarize the difficulties below.

Firstly, although there are some important progresses \cite{Dong,Solodukhin1,Camps},  due to the 'splitting problem' \cite{Miao1,Miao2}, the exact entropy formula for higher curvature gravity is still unknown.

Secondly, we do not know where the entangling surface is located for higher curvature gravity. In other words, we do not know on which surface to apply the entropy formula. There are two methods to determine the location of the entangling surface in the bulk. The first one is the so-called 'boundary condition method': one require that equations of motion are regular on the entangling surface \cite{Maldacena1}. This method can yield the correct entangling surfaces for Einstein gravity and Lovelock gravity \cite{Dong,Maldacena1,Camps1}. However, so far it is not clear whether this approach can give reasonable results for general higher curvature gravity. The second method is the so-called 'cosmic brane method'. One takes the variation of the entropy functional and identifies the entangling surface with the extremal entropy surface. However, there are more than one extremal entropy surfaces in higher curvature gravity. One need extra conditions to fix the arbitrariness \cite{Erdmenger}.

Thirdly, there are infinite parameters in general higher curvature gravity. If we study them case by case, then it is impossible to give a general proof. 

Our resolutions to the above difficulties are as follows. Let us discuss them one by one. Firstly, we do not need the exact entropy formula of higher curvature gravity for the proof of the conjecture (\ref{conjecture}). Take into account the 'splitting problem', the correct entropy formula in $AdS_4$ differs from the one proposed in \cite{Dong} by some higher extrinsic curvature terms $K^{2m}$ with $m\ge 2$. It turns out that these higher extrinsic curvature terms do not affect either $\sigma$ or $C_T$. Secondly, as argued in \cite{Myers0,Myers1}, it is natural to work in the perturbative framework in order to suppress the massive modes and ghost modes. In the perturbative framework, the extremal entropy surface is unique and well-defined. It is sightly deformed away from the minimal surface and can yield correct universal terms of EE for even-dimensional CFTs. Thirdly, we use the 'background field approach' developed for the holographic Weyl anomaly and entanglement entropy \cite{Miao}. We expand the action around a background curvature. It turns out that only the first few terms in the expansions contribute to the universal terms of entanglement entropy. Thus, we only need to deal with finite rather than infinite terms. 

In addition to the holographic proof of the conjecture (\ref{conjecture}) \cite{Myers0}, we also clarify some interesting problems discussed in \cite{Myers0,Myers1}. Firstly, we find that, in contrast to $\sigma/C_T$, $\kappa/C_T$ is not a universal ratio. In other words, the behaviour of logarithmic terms of EE is not universal in the singular limit. Secondly, we notice that the lower bound $a_E(\Omega)/C_T$ associated to Einstein gravity can be violated by general higher curvature gravity. 

By studying the holographic models, we find that there are similar universal laws for CFTs in higher dimensions. For simplicity, we focus on the singularities from the higher-dimensional cones. The corresponding bulk metric takes the form
\begin{eqnarray}\label{Hcones}
ds^2=\frac{dz^2+dt^2_E+d\rho^2+\rho^2(d\theta^2+\sin^2\theta d\Omega^2_{d-3})}{z^2}
\end{eqnarray}
where $d\Omega^2_{d-3}$ is the metric of unit $(d-3)$-sphere. We have $\theta\in [0, \Omega]$ with $\Omega \le \pi$. The universal terms of EE are logarithmic terms $-a_d(\Omega)\log(H/\delta)$ and squared logarithmic terms $-a_d(\Omega)\log^2(H/\delta)$ in odd dimensions and even dimensions, respectively. Similar to the 3d CFTs, we have $a_d(\Omega)=a_d(\pi-\Omega)$ and the following asymptotic behaviors
\begin{eqnarray}\label{Hconesbe}
a_d(\Omega\to \pi/2)\simeq \sigma_d (\pi/2-\Omega)^2,\ \ \ a_d(\Omega\to 0)\simeq \kappa_d/\Omega
\end{eqnarray}
Based on holographic results, we conjecture that
\begin{eqnarray}\label{conjecture2}
\sigma_d/C_T=\sigma_{d,E}/C_{T,E}=c_d
\end{eqnarray}
are universal ratios for general CFTs. Here $C_T$ are the central charges defined in eq.(\ref{twopoint}), 'E' denotes Einstein gravity and $c_d$ are universal constants which only depend on the dimensions. 

The paper is organized as follows. In Sect.2, we study the central charge $C_T$ for CFTs dual to general higher curvature gravity. In Sect.3, we briefly review the holographic entanglement entropy and derive a formal entropy formula for general higher curvature gravity in $AdS$. In Sect.4, we give a holographic proof of the conjecture of \cite{Myers0}. In Sect.5, we find similar univeral laws for CFTs in higher dimensions. Finally, we conclude in Sect.6.

Note added: While we were finishing this paper, the work \cite{Alishahiha} appeared in arXiv and it seems to have some overlaps in the universal ratios for CFTs in higher dimensions. It should be mentioned that \cite{Myers0,Myers1} have also anticipated the generalizations of the universal ratios to higher dimensions. Later they derive a nice formula of the universal ratios in general dimensions in \cite{Myers5}.

\section{The holographic central charges}

In this section, we discuss the central charge $C_T$ for CFTs dual to general higher derivative gravity $f(R_{\mu\nu\sigma\rho})$. We obtain a very simple expression for $C_T$ and find that $C_T$ is the coefficient of the Weyl-squared term in the Weyl anomaly. For examples, $C_T$ is the 'c' charge relevant to the $C^2$ term in the 4d Weyl anomaly
 \begin{eqnarray}\label{4danomaly1}
 <T^i_i>=\frac{c}{16\pi^2}C_{ijkl}C^{ijkl}-\frac{a}{16\pi^2}E_4.
 \end{eqnarray}
And $C_T$ is the '$B_3$' charge relevant to the $C\Box C$ term in the 6d Weyl anomaly
 \begin{eqnarray}\label{6danomaly1}
 <T^i_i>=\sum_{n=1}^{3}B_n I_n +2 A E_6
 \end{eqnarray}
where $I_3\sim C_{ijkl}\Box C^{ijkl}+... $.

We use the 'background field approach' introduced in \cite{Miao}. This method together with \cite{Theisen1,Theisen2} are very useful tools to derive the holographic Weyl anomaly and universal terms of EE \cite{Miao2,Miao}. Firstly, we define a 'background-curvature' (we set the AdS radius $\tilde{H}=1$ below)
\begin{eqnarray}\label{backcurvature}
\tilde{R}_{\mu\nu\sigma\rho}=g_{\mu\rho}g_{\nu\sigma}-g_{\mu\sigma}g_{\nu\rho}
\end{eqnarray}
and denote the difference between the curvature and the  'background-curvature' by
\begin{eqnarray}\label{diffcurvature}
\bar{R}_{\mu\nu\sigma\rho}=R_{\mu\nu\sigma\rho}-\tilde{R}_{\mu\nu\sigma\rho}.
\end{eqnarray}
Then we expand the action around this 'background-curvature' and get \cite{Miao}
\begin{eqnarray}\label{GHCaction0}
I&=&\frac{1}{16\pi G}\int d^{d+1}x\sqrt{g} f(R_{\mu\nu\sigma\rho})\nonumber\\
&=&\frac{1}{16\pi G}\int d^{d+1}x\sqrt{g}\big[ f_0+ c^{(1)}_1 \bar{R} +( c^{(2)}_1\bar{R}_{\mu\nu\sigma\rho}\bar{R}^{\mu\nu\sigma\rho}+ c^{(2)}_2\bar{R}_{\mu\nu}\bar{R}^{\mu\nu}+ c^{(2)}_3\bar{R}^2)+O(\bar{R}^3) \big]
\end{eqnarray}
where $f_0=f(\tilde{R}_{\mu\nu\sigma\rho})=f(R_{\mu\nu\sigma\rho})|_{AdS}$ is the Lagrangian for pure AdS, $c^n_i$ are some constants determined by the action. We require that the higher derivative gravity has an asymptotic AdS solution. This would impose a condition $ c^{(1)}_1=-f_0/2d$ \cite{Miao}. Using this condition, we can rewrite the action (\ref{GHCaction0}) as
\begin{eqnarray}\label{GHCaction1}
I=\frac{1}{16\pi G}\int d^{d+1}x\sqrt{g}\big[-\frac{f_0}{2d}(R+d^2-d) +( c^{(2)}_1\bar{R}_{\mu\nu\sigma\rho}\bar{R}^{\mu\nu\sigma\rho}+ c^{(2)}_2\bar{R}_{\mu\nu}\bar{R}^{\mu\nu}+ c^{(2)}_3\bar{R}^2)+O(\bar{R}^3) \big]
\end{eqnarray}

Following \cite{Myers0,Myers1}, we consider small metric fluctuations in the AdS background. Imposing the transverse traceless gauge  $\bar{\nabla}^{\mu}h_{\mu\nu}=0$ and $g^{\mu\nu}h_{\mu\nu}=0$, we can derive the linearized equations of motion for higher curvature gravity (\ref{GHCaction1}) as
\begin{eqnarray}\label{EOMHDG1}
-\frac{\alpha_d}{2}[\bar{\Box}+\frac{2}{\bar{L}^2}]h_{\mu\nu}-\frac{ c^{(2)}_2+4 c^{(2)}_1}{2}[\bar{\Box}+\frac{2}{\bar{L}^2}]^2h_{\mu\nu}=8\pi G T_{\mu\nu}
\end{eqnarray}
where $\alpha_d$ is given by
\begin{eqnarray}\label{alphad}
\alpha_d=-\frac{f_0}{2d}+(4d-8) c^{(2)}_1.
\end{eqnarray}
Remarkably, $\alpha_d$ only depends on two parameters of the general higher curvature gravity (\ref{GHCaction0}). 
Note that the second term of eq.(\ref{EOMHDG1}) is highly suppressed near the physical pole, i.e. $[\bar{\Box}+\frac{2}{\bar{L}^2}]h_{\mu\nu}\simeq 0$. Comparing eq.(\ref{EOMHDG1}) with the linearized Einstein equations eq.(\ref{EOME}), we find that the effective Newton constant of the general higher curvature gravity (\ref{GHCaction0}) is $G_{eff}=G/\alpha_d$. Note that the central charge $C_{T,E}$ for CFTs dual to Einstein gravity is 
\begin{eqnarray}\label{CTE}
C_{T,E}=\frac{d+1}{d-1}\frac{\Gamma[d+1]}{\pi^{d/2}\Gamma[d/2]}\frac{\tilde{L}^{d-1}}{8\pi G}.
\end{eqnarray}
 Thus we get
\begin{eqnarray}\label{CTGHCG}
C_T=\alpha_d C_{T,E}
\end{eqnarray}
Let us comment the above results.

Firstly, with the help of the expansions (\ref{EOMHDG1}) we get a very  simple and general expression of $\alpha_d$ (\ref{alphad}) for higher curvature gravity. It agrees with eq.(\ref{alpha0}) for the holographic model (\ref{HDG0}). It is much simpler and thus enables us to discuss the higher curvature gravity generally rather than to study them case by case. 

Secondly, thanks to the simplicity of $\alpha_d$ (\ref{alphad}), the physical meaning of $C_T$ (\ref{CTGHCG}) becomes clear. It is the central charge related to the Weyl-squared term in the Weyl anomaly. For example, $C_T$ is proportional to the c charge in the 4d Weyl anomaly
\begin{eqnarray}\label{4danomaly}
<T^i_i>=\frac{c}{16\pi^2}C_{ijkl}C^{ijkl}-\frac{a}{16\pi^2}E_4.
\end{eqnarray}
Here $c=\frac{\pi}{8G}(-\frac{f_0}{8}+8 c^{(2)}_1)\sim C_T$ \cite{Miao}. As another example,  $C_T$ is proportional to the $B_3$ charge in the 6d Weyl anomaly
\begin{eqnarray}\label{6danomaly}
<T^i_i>=\sum_{n=1}^{3}B_n I_n +2 A E_6
\end{eqnarray}
where $I_3\sim C_{ijkl}\Box C^{ijkl}+... $ and $B_3=\frac{1}{3072\pi G}(-\frac{f_0}{12}+16 c^{(2)}_1)\sim C_T$ \cite{Miao}.

Thirdly, due to the fact $ O(\bar{R}^3)\sim O(h^3)$, the $O(\bar{R}^3)$ terms in the action (\ref{GHCaction1}) do not affect the linearized E.O.M. As a result, $\alpha_d$ and thus $C_T$ are independent of such terms. To prove the conjecture $\sigma / C_T=\pi^2/24$ for 3d CFTs, we need to prove that the entropy from the  $O(\bar{R}^3)$ terms does not contribute to $\sigma$. 

\section{The holographic entanglement entropy}

In this section, we derive a formal entropy formula for the general higher curvature gravity. We utilize this formula to prove the conjecture eq.(\ref{conjecture}) in the next section. Now let us briefly review the derivations of the holographic entanglement entropy for higher curvature gravity \cite{Dong}. We start with the regularized conical metric in a coordinate system adapted to a neighborhood of the conical singularity \cite{Dong}:
\begin{eqnarray}\label{conemet}
ds^2 = e^{2A} \[dz d\zb + e^{2A} T (\zb dz-z d\zb)^2 \] + \(\gamma_{ij} + 2K_{aij} x^a + Q_{abij} x^a x^b\) dy^i dy^j \nonumber\\
+ 2i e^{2A} U_i \(\zb dz-z d\zb\) dy^i + \cdots \,.
\end{eqnarray}
Here $x^a \in \{z,\zb\}$ denotes orthogonal directions to the conical singularity, and $y^i$ denotes parallel directions.  The regularized warp factor is
\be
A = -\fr{\e}{2} \log (z\zb+b^2) \,, \qqu
\e \eq 1-\fr{1}{n} \,,
\ee
Using the replica trick, one can derive the entropy as 
\begin{eqnarray}\label{HEE}
S=-\partial_{\epsilon} I_{reg}|_{\epsilon=0}
\end{eqnarray} 
where $I_{reg}$ is the gravitational action got from the regularized metric (\ref{conemet}). There are two kinds of terms relevant to the entropy. The first kind is 
\begin{eqnarray}\label{dA}
&&R_{z\bar{z}z\bar{z}}=e^{2A}\partial_z\partial_{\bar{z}}A+...\nonumber\\
&&\int dz d\bar{z} \partial_z\partial_{\bar{z}}A=-\pi \epsilon.
\end{eqnarray}
It contributes to Wald entropy. The second kind is 
\begin{eqnarray}\label{dAdA}
&&R_{zizj}=2K_{zij}\partial_zA+..., \ R_{\bar{z}k\bar{z}l}=2K_{\bar{z}kl}\partial_{\bar{z}}A+...\nonumber\\
&&\ \ \ \ \ \ \ \ \ \ \ \ \ \ \int dz d\bar{z} \partial_zA\partial_{\bar{z}}Ae^{-\beta A}=-\frac{\pi \epsilon}{\beta} .
\end{eqnarray}
This is the would-be logarithmic term and it contributes to the anomaly-like entropy \cite{Dong}. 

Applying eqs.(\ref{HEE},\ref{dA},\ref{dAdA}), one can derive the holographic entanglement entropy (HEE) for general higher curvature gravity $f(R_{\mu\nu\sigma\rho})$ \cite{Dong}
\begin{eqnarray}\label{HEEDong}
S_{\text{HEE}}&=&\frac{1}{8 G}\int d^{d-1}y \sqrt{\gamma}\big{[}\ \frac{\partial f}{\partial R_{z\bar{z}z\bar{z}}}
+16\sum_{\beta}(\frac{\partial^2 f}{\partial R_{zizj}\partial R_{\bar{z}k\bar{z}l}})_{\beta}\frac{K_{zij}K_{\bar{z}kl}}{\beta+2} \ \big{]}
\end{eqnarray}
Here $\beta$ come from the formula (\ref{dAdA}), and $(\frac{\partial^2 f}{\partial R_{zizj}\partial R_{\bar{z}k\bar{z}l}})_{\beta}$ are the coefficients in the expansions
\begin{eqnarray}\label{HEEbeta}
\frac{\partial^2 f}{\partial R_{zizj}\partial R_{\bar{z}k\bar{z}l}}=\sum_{\beta}e^{-\beta A}(\frac{\partial^2 f}{\partial R_{zizj}\partial R_{\bar{z}k\bar{z}l}})_{\beta}.
\end{eqnarray}

 \cite{Dong} proposes to regularize $Q_{z\bar{z}ij}$ as $e^{2A}Q_{z\bar{z}ij}$. Later it is found that this ansatz yields inconsistent results for the universal terms of EE for 6d CFTs \cite{Miao2}. To resolve this inconsistency, \cite{Miao2} proposes the following regularizations
 \begin{eqnarray}\label{TVQ1}
 &&T=e^{-2A}T_0+T_{1},\nonumber\\
 &&Q_{\ z \bar{z} ij}=Q_{0\ \ z \bar{z}ij}+e^{2A}Q_{1\ z\bar{z}ij}
 \end{eqnarray}
 How to split $M$ into $M_0$ and $M_1$ ($M$ denotes $T$ and $Q$) is the so-called the splitting problem. It appears because one cannot distinguish $r^2$ and $r^{2n}$ in the expansions of the conical metric. It is expected that the splitting problem can be fixed by using E.O.M. This is indeed the case for Einstein gravity \cite{Miao1,Miao2}. However, it is a highly non-trivial problem to fix the splittings for general higher curvature gravity. Without resolution to this problem, we cannot apply the formula (\ref{HEEDong}) to derive the entropy. 
 
 To be consistent with Wald entropy on entangling surface with rotational symmetry, $T_0$ and $Q_0$ must be functions of the extrinsic curvatures, i.e., $T_0\sim K^2$ and  $Q_0\sim K^2$ \cite{Miao1,Miao2} (This is indeed the case for the splittings obtained from Einstein equations.) As a result, the correct entropy may differ from the original one proposed by \cite{Dong} by some $O(K^4)$ terms. It turns out that these $O(K^4)$ terms do not contribute to the universal terms of EE for 4d CFTs on smooth entangling surfaces \cite{Miao}. Although the story is a little different for universal terms of EE on singular entangling surfaces, the $O(K^4)$ terms are still less important if we focus on the near smooth region. As we shall prove in the next section, the coefficient $\sigma$ defined in the smooth limit eq.(\ref{acondition1}) is indeed independent of these $O(K^4)$ terms. 
 
 Take into account the splittings, the Riemann tensors near the  the conical singularity $z \to 0$ take the form
  \begin{eqnarray}\label{Riemanntensor}
  R_{\mu\nu\sigma\rho}&=&e^{2[P/2]A}  R_{1 \mu\nu\sigma\rho}+ e^{2([P/2]-1)A}  (K^2)_{\mu\nu\sigma\rho}\nonumber\\
  &=& e^{2[P/2]A}  R_{\mu\nu\sigma\rho}|_{A=0}+ (e^{2([P/2]-1)A}-e^{2[P/2]A}) (K^2)_{\mu\nu\sigma\rho}
  \end{eqnarray}
where $P$ are the numbers of $(z,\bar{z})$ appearing in ($\mu\nu\sigma\rho$), $(K^2)_{\mu\nu\sigma\rho}$ denotes some extrinsic curvature squared terms, $ R_{\mu\nu\sigma\rho}|_{A=0}$ are the Riemann tensors without regularization. If we allow more general splittings for the conical metrics, i.e., we expand $T, U, Q$ of (\ref{conemet}) in infinite powers of $e^{-2A}$. Then the Riemann tensors (\ref{Riemanntensor}) can be generalized as
  \begin{eqnarray}\label{Riemanntensor1}
  R_{\mu\nu\sigma\rho}= e^{2[P/2]A}  R_{\mu\nu\sigma\rho}|_{A=0}+\sum_{i=1}^{\infty} e^{2[P/2]A}(e^{-2i A}-1) (K^2)_{i\ \mu\nu\sigma\rho}
  \end{eqnarray}
  For pure AdS, it can be further simplifed as
 \begin{eqnarray}\label{Riemanntensor2}
 R_{\mu\nu\sigma\rho}=g_{\mu\rho}g_{\nu\sigma}-g_{\mu\sigma}g_{\nu\rho} +\sum_{i=1}^{\infty} e^{2[P/2]A}(e^{-2i A}-1) (K^2)_{i\ \mu\nu\sigma\rho}
 \end{eqnarray} 
 
Now we are ready to derive the entropy for general higher curvature gravity (\ref{GHCaction1}). For pure AdS, using eqs.(\ref{HEEDong},\ref{Riemanntensor2}), we obtain
\begin{eqnarray}\label{HEEMiao}
S=-\frac{1}{4 G}\int d^{d-1}y \sqrt{\gamma}\big{[} -\frac{f_0}{2d}- c^{(2)}_2\frac{1}{2}(tr K)^2-2  c^{(2)}_1 tr K^2 +\sum_{m=2}^{\infty} \lambda_{m} (K^{2m})  \big{]}
\end{eqnarray} 
 where $(K^{2m})$ denote all the possible higher extrinsic curvature terms of order $O(K^{2m})$ and $\lambda_{m}$ are some constants related the higher curvature gravity. Note that eq.(\ref{HEEMiao}) works in the Euclidean signature, which differs from its Lorentzian form by a minus sign. 

\section{A holographic proof of the conjecture } 

In this section, we firstly briefly review the corner contributions to holographic entanglement entropy for Einstein gravity, and then give a holographic proof of the conjecture eq.(\ref{conjecture}). We focus on pure $AdS_4$ and work in the perturbative framework for higher curvature gravity.

\subsection{Einstein gravity}

In this subsection, we review the corner contributions to holographic entanglement entropy for Einstein gravity \cite{Hirata,Myers1}. Let us start with the Euclidean $AdS_4$ in Poincare coordinates
\begin{eqnarray}\label{AdS4}
ds^2=\bar{L}^2\frac{dz^2+dt^2_E+d\rho^2+\rho^2 d\theta^2}{z^2},
\end{eqnarray}
where $\bar{L}$ is the radius of $AdS_4$ and we set it to 1 below. The holographic entanglement entropy for Einstein gravity is given by \cite{Ryu1,Ryu2}
\begin{eqnarray}\label{HEEEinstein}
S_{\text{EE}}=\frac{1}{4 G}\int_{\Sigma} d^{2}y \sqrt{\gamma}
\end{eqnarray}
where $\Sigma$ is the bulk minimal surface which is homologous to the entangling surface on the boundary. Let us take  $t_E=0, \theta\in [-\Omega/2, \Omega/2]$ to denote the entangling surface on the boundary. It has a sharp corner for $ \Omega \not=\pi$. The bulk minimal surface can be parametrized as $z=z(\rho,\theta)$. Take into account the scaling symmetry of $AdS$, we can simplify the ansatz to $z=\rho h(\theta)$. With this ansatz, the entanglement entropy (\ref{HEEEinstein}) becomes
\begin{eqnarray}\label{HEEEinstein1}
S_{\text{EE}}=\frac{1}{2G}\int_{\delta/h_0}^{H}\frac{d \rho}{\rho}\int_{0}^{\Omega/2-\epsilon}d\theta \frac{\sqrt{1+h^2+(h')^2}}{h^2}
\end{eqnarray}
where $H$ is the size of the entangling surface on the boundary, $\delta$ is the cutoff for $z$, $h_0=h(0)$ and the angular cut-off $\epsilon$ is defined at $z=\delta$, i.e., $\rho\  h(\Omega/2-\epsilon)=\delta$.  

Note that there is no explicit $\theta$ dependence in eq.(\ref{HEEEinstein1}), thus the corresponding 'Hamiltonian' is a conserved quantity. We get
\begin{eqnarray}\label{hsolutionEin}
\frac{1+h^2}{h^2\sqrt{1+h^2+(h')^2}}=\frac{\sqrt{1+h_0^2}}{h_0^2}
\end{eqnarray} 
where we have used $h'_0=0$ from symmetry.
Using eq.(\ref{hsolutionEin}), we can rewrite eq.(\ref{HEEEinstein1}) as
\begin{eqnarray}\label{HEEEinstein2}
S_{\text{EE}}=\frac{1}{2G}\int_{\delta/h_0}^{H}\frac{d \rho}{\rho}\int_{0}^{\sqrt{(\rho/\delta)^2-1/h_0^2}}dy\sqrt{\frac{1+h_0^2(1+y^2)}{2+h_0^2(1+y^2)}}
\end{eqnarray}
with $y=\sqrt{1/h^2-1/h_0^2}$. 

Let us focus on the universal logarithmic divergence of EE. From eq.(\ref{HEEEinstein2}), it is easy to observe that the logarithmic divergence comes from the integral $d\rho$. To get the coefficient of the logarithmic divergence, we need to extract the finite part from the integral $dy$. Take into account the boundary behaves $(y\to \infty)$ 
\begin{eqnarray}\label{integral0}
\sqrt{\frac{1+h_0^2(1+y^2)}{2+h_0^2(1+y^2)}}\sim 1+O(1/y^2),
\end{eqnarray}
we can derive the logarithmic divergence of EE as
\begin{eqnarray}\label{HEEEinstein3}
S_{\text{EE, log}}=-a(\Omega)\log(\frac{H}{\delta})
\end{eqnarray}
where $a(\Omega)$ is given by
\begin{eqnarray}\label{aEin}
a(\Omega)=\frac{1}{2G}\int_{0}^{\infty}dy [1-\sqrt{\frac{1+h_0^2(1+y^2)}{2+h_0^2(1+y^2)}}]
\end{eqnarray}
And the opening angle $\Omega$ can be obtained from 
\begin{eqnarray}\label{angleEin}
\Omega=2\int_{0}^{h_0}dh \frac{h^2\sqrt{1+h_0^2}}{\sqrt{1+h^2}\sqrt{(h_0^2-h^2)(h_0^2+(1+h_0^2)h^2)}}
\end{eqnarray}
For small opening angle ($\Omega\to 0$), we find 
\begin{eqnarray}\label{smallangle0}
\Omega&=&\frac{2\sqrt{\pi}\Gamma(\frac{3}{4})}{\Gamma(\frac{1}{4})}h_0+O(h_0^3)\\
a_E(\Omega)&=&\frac{1}{2G}\Gamma(\frac{3}{4})^4\frac{1}{\Omega}+O(\Omega)\label{smalla0}
\end{eqnarray}
For the near-smooth case ($\Omega\to \pi$), we have 
\begin{eqnarray}\label{largeangle0}
\pi-\Omega&=&\frac{\pi}{h_0}+O(\frac{1}{h_0^3})\\
a_E(\Omega)&=&\frac{1}{8\pi G}(\pi-\Omega)^2+O(\pi-\Omega)^4\label{largeangleaE0}
\end{eqnarray}
From the above equations, we get $\kappa_E=\frac{1}{2G}\Gamma(\frac{3}{4})^4$ and $\sigma_E=\frac{1}{8\pi G}$. Recall that $C_{T,E}=\frac{3}{\pi^3 G}$, we obtain
\begin{eqnarray}\label{Einsteinresult1}
\frac{\sigma_E}{C_{T,E}}=\pi^2/24
\end{eqnarray}
which agrees with the conjecture (\ref{conjecture}). In addition to eq.(\ref{Einsteinresult1}), we also have
\begin{eqnarray}\label{Einsteinresult2}
\frac{\kappa_E}{C_{T,E}}=\frac{\pi^2}{6}\Gamma(\frac{3}{4})^4
\end{eqnarray}
which seems to be another universal law from the holographic study of \cite{Myers1}. However, as we shall show in the next section, this is not the case for general higher curvature gravity.

\subsection{General higher curvature gravity}

In this subsection, we investigate the corner contribution to EE for CFTs dual to higher curvature gravity. Let us start with the holographic entanglement entropy for the general higher curvature gravity $f(R_{\mu\nu\sigma\rho})$ in $AdS_4$ (\ref{HEEMiao})
\begin{eqnarray}\label{HEEMiao3d}
S=\frac{1}{4 G}\int_{\Sigma} d^{2}y \sqrt{\gamma}\big{[} -\frac{f_0}{6}- c^{(2)}_2\frac{1}{2}(tr K)^2-2  c^{(2)}_1 tr K^2 +\sum_{m=2}^{\infty} \lambda_{m} (K^{2m})  \big{]}
\end{eqnarray}
where $\Sigma$ denotes the extremal entropy surface, $f_0,  c^{(2)}_1,  c^{(2)}_2, \lambda_m$ are the parameters of the higher curvature gravity, and $(K^{2m})$ denote all the possible higher extrinsic curvature terms of order $O(K^{2m})$. Note that eq.(\ref{HEEMiao3d}) works in the Lorentzian signature which differs from its Euclidean expression (\ref{HEEMiao}) by a minus sign. Following \cite{Myers0,Myers1}, we work in the perturbative framework with $( c^{(2)}_1,  c^{(2)}_2, \lambda_m \ll 1 )$ in order to suppress the massive modes and ghost modes in higher gravity gravity.

Let us firstly discuss the squared extrinsic curvature terms $O(K^2)$ in eq.(\ref{HEEMiao3d}). Because $tr K=0$ on the extremal area surface, the minimal surface will also extremize the entropy functional $\sqrt{\gamma}(tr K)^{2n}$ ($n\ge 1$). Thus, we can drop such terms in eq.(\ref{HEEMiao3d}) in the perturbative framework. Using the Gauss-Codazzi equations in $AdS_4$, we can rewrite $\sqrt{\gamma} Tr K^2$ as
\begin{eqnarray}\label{trKK}
\int_{\Sigma} d^{2}y \sqrt{\gamma}Tr K^2=\int_{\Sigma} d^{2}y \sqrt{\gamma}(-2-\mathcal{R}+(tr K)^2)=-2\int_{\Sigma} d^{2}y \sqrt{\gamma}
\end{eqnarray}
where  $\mathcal{R}$ is the intrinsic curvature, and we have dropped $(tr K)^2$ and a total derivative $\sqrt{\gamma}\mathcal{R}$ in the above equation. Take into account eq.(\ref{trKK}) and $(trK)^2\sim 0$, eq.(\ref{HEEMiao3d}) becomes
\begin{eqnarray}\label{HEEMiao3d1}
S=\frac{1}{4 G}\int_{\Sigma} d^{2}y \sqrt{\gamma}\big{[} -\frac{f_0}{6}+4  c^{(2)}_1 +\sum_{m=2}^{\infty} \lambda_{m} (K^{2m})  \big{]}
\end{eqnarray}

Let us go on to study the higher extrinsic curvature terms $O(K^{2m})$ with $m\ge 2$. Since $\lambda_m\ll 1$, we focus on the leading order of $\lambda_m$ below. It turns out that, at order $O(\lambda_m)$, all the possible terms of $(K^{2m})$ are either zero or equivalent to $Tr K^{2m}$ up to some overall factor. Note that there are only two eigenvalues $k_{\pm}$ for the extrinsic curvature $K^i_j$ in $AdS_4$. Then, similar to $(Tr K=k_++k_-=0)$, the trace of odd powers of the extrinsic curvature
\begin{eqnarray}\label{TrKodd}
Tr K^{2l-1}=k_+^{2l-1}+k_-^{2l-1}=0
\end{eqnarray}
vanishes on extremal area surfaces. As a result, we can drop all the terms including the trace of odd powers of the extrinsic curvature. That is because there are at leat two 'odd-trace' terms in $(K^{2m})$, and similar to $(Tr K)^2$, such terms neither change the action or  E.O.M at the leading order of $O(\lambda_m)$. 

Now let us discuss the terms including only the trace of even powers of the extrinsic curvature, i.e., $\prod_{i=1}^{n}Tr K^{2m_i}$ with $\sum_{i=1}^{n\le m}m_i=m$. As we shall show below, it is equivalent to $2^{n-1}Tr K^{2m}$. To see this, let us check the action and E.O.M at order $O(\lambda_m)$ below.

For the action, we have 
\begin{eqnarray}\label{checkaction}
\lambda_m \prod_{i=1}^{n}Tr K^{2m_i}=\lambda_m 2^n k_+^{2m}+O(\lambda_m^2)=\lambda_m 2^{n-1} Tr K^{2m}+O(\lambda_m^2)
\end{eqnarray}

For E.O.M, it is equivalently to consider the variation of the action
\begin{eqnarray}\label{checkEOM}
\lambda_m \delta (\prod_{i=1}^{n}Tr K^{2m_i})=\lambda_m  2^{n}m k_+^{2m-1}(\delta k_+-\delta k_-)+O(\lambda_m^2)=\lambda_m \delta(2^{n-1} Tr K^{2m}) +O(\lambda_m^2)
\end{eqnarray}
Now it is clear that $\prod_{i=1}^{n}Tr K^{2m_i}$ and $2^{n-1}Tr K^{2m}$ yield the same action and E.O.M at the first order of $O(\lambda_m)$.
Thus, we can label all the possible terms of $(K^{2m})$ by one term $Tr K^{2m}$. Then the entropy eq.(\ref{HEEMiao3d1}) becomes
\begin{eqnarray}\label{HEEMiao3d2}
S=\frac{-\frac{f_0}{6}+4  c^{(2)}_1}{4 G}\int_{\Sigma} d^{2}y \sqrt{\gamma}\big{[} 1 +\sum_{m=2}^{\infty} \bar{\lambda}_{m} Tr K^{2m}  \big{]}+O(\lambda_{m})^2
\end{eqnarray}
where we have rescaled $\lambda_{m}= (-\frac{f_0}{6}+4  c^{(2)}_1)\bar{\lambda}_{m}$. If the higher extrinsic curvature terms $Tr K^{2m}$ vanish, from eq.(\ref{HEEMiao3d2}) we can easily obtain 
\begin{eqnarray}\label{sigmaHCG}
\sigma=(-\frac{f_0}{6}+4  c^{(2)}_1)\sigma_E.
\end{eqnarray}
Recall that $C_T=(-\frac{f_0}{6}+4  c^{(2)}_1)C_{T,E}$ eq.(\ref{CTGHCG}) for general higher curvature gravity, we get
\begin{eqnarray}\label{conjectHCG}
\frac{\sigma}{C_T}=\frac{\sigma_E}{C_{T,E}}=\frac{\pi^2}{24}.
\end{eqnarray}
Thus, to prove the conjecture (\ref{conjecture}), we need and only need to prove that the higher extrinsic curvature terms $Tr K^{2m}$ do not contribute to $\sigma$.

To proceed, let us derive the exact expression of the entropy eq.(\ref{HEEMiao3d2})
\begin{eqnarray}\label{HEEMiao3d3}
S&=&\frac{-\frac{f_0}{6}+4  c^{(2)}_1}{2G}\int_{\delta/h_0}^{H}\frac{d \rho}{\rho}\int_{0}^{\Omega/2-\epsilon}d\theta \frac{\sqrt{1+h^2+h'^2}}{h^2}\big[1\nonumber\\
&&\ \ +\sum_{m=2}^{\infty} \bar{\lambda}_{m} \big((\frac{1}{\sqrt{h'^2+h^2+1}})^{2m}+(\frac{h^4+h h''+2 h^2+\left(h'\right)^2+h^3 h''+1}{(h^2+\left(h'\right)^2+1)^{3/2}})^{2m}\big)\big]
\end{eqnarray}
Similar to the case of Einstein gravity, there is no explicit $\theta$ dependence in eq.(\ref{HEEMiao3d3}). Thus we can derive a first integral
\begin{eqnarray}\label{firstintegral}
\frac{h^2+1}{h^2 \sqrt{h^2+\left(h'\right)^2+1}}+\sum_{m=2}^{\infty} \bar{\lambda}_{m}F_m[h,h',h'',h^{(3)}]=\frac{\sqrt{h_0^2+1}}{h_0^2}+\sum_{m=2}^{\infty} \bar{\lambda}_{m}F_m[h_0,0,h''_0,0]
\end{eqnarray}
where $F_m[h,h',h'',h^{(3)}]$ is a very complicated function given by eq.(\ref{FMiao}) in the appendix. Let us solve eq.(\ref{firstintegral}) perturbatively. After some algebra, we get
\begin{eqnarray}\label{H1Miao}
h'&=&-\frac{\sqrt{\left(h^2+1\right) \left(h^4 \left(-\left(h_0^2+1\right)\right)+h^2 h_0^4+h_0^4\right)}}{h^2 \sqrt{h_0^2+1}}\nonumber\\
&+&\sum_{m=2}^{\infty}\bar{\lambda}_{m} \frac{2\left(h^2+1\right)^{\frac{3}{2}-2 m} h^{4 m-6} h_0^{4-4 m}}{\sqrt{\left(h^2+1\right) h_0^4-h^4 \left(h_0^2+1\right)} \left(h_0^2+1\right)^{\frac{3}{2}-m}}\bigg(-\frac{\left(h^2+1\right)^{2 m} \left(\left(h_0^2+2\right) m-1\right)}{h^{4 m-4} h_0^{-4 m} \left(h_0^2+1\right)^{2 m-1}} \nonumber\\
&&+\left(h^4 \left(h_0^4 m+h_0^2 \left(-4 m^2+5 m-1\right)-4 m^2+5 m-1\right)+2 h^2 h_0^4 m (2 m-1)+h_0^4 m (4 m-3)\right)\bigg)\nonumber\\
&+&O(\bar{\lambda}_{m})^2\\
h''&=&-\frac{h^6 \left(h_0^2+1\right)+2 h^2 h_0^4+2 h_0^4}{h^5 \left(h_0^2+1\right)}+O(\bar{\lambda}_{m})\label{H2Miao}
\end{eqnarray}
Using eq.(\ref{H1Miao}), we can express the opening angle $\Omega$ in the function of $h_0$
\begin{eqnarray}\label{angleHCG}
\Omega=2\int_{h_0}^{0}dh\frac{1}{h'}=\Omega_E(h_0)+\sum_{m=2}^{\infty}\bar{\lambda}_{m} \Omega_m(h_0)+O(\bar{\lambda}_{m})^2
\end{eqnarray}
where $\Omega_E(h_0)$ is the function eq.(\ref{angleEin}) for Einstein gravity, and it behaves as $\Omega_E(h_0)\sim \pi-\frac{\pi}{h_0}$ for large $h_0$. In the smooth limit $h_0\to \infty$, we can derive $\Omega_m$ as
\begin{eqnarray}\label{angleHCGm}
\Omega_m&=&-\frac{1}{h_0^{2 m-1}}\int_{0}^{\infty}dy\frac{4 m \left(\frac{\left(y^2+1\right)^{2m}-1}{y^2}+(3-4 m) y^2+(2-4 m) \right)}{\left(y^2+1\right)^{2 m+1}}+O(\frac{1}{h_0^{2m+1}})\nonumber\\
&=&  \frac{\omega_m\pi}{h_0^{2m-1}}+O(\frac{1}{h_0^{2m+1}})
\end{eqnarray}
where we have substituted $y=\sqrt{1/h^2-1/h_0^2}$. And $\omega_m$ are some finite numbers given by
\begin{eqnarray}\label{omegam}
\omega_1=0,\ \omega_2=3/2,\ \omega_3=195/64,\ \omega_4=595/128,\ \omega_5=103065/16384,...
\end{eqnarray}
 Remarkably, $\Omega_m\sim 1/h_0^{2m-1}$ with $m\ge 2$ are the subleading corrections to the angle function. Note that we have $\omega_1=0$, which means that the extrinsic curvature squared term $ Tr K^2 $ does not modify the angle function. That is reasonable. Recall that $\sqrt{\gamma} Tr K^2$ is equivalent to $-2\sqrt{\gamma} $, thus the angle function for $\sqrt{\gamma} Tr K^2$ should be exactly the same as that for Einstein gravity. This can be regarded as a check of our results. 
 
 Now let us study the logarithmic term of EE, i.e., $-a(\Omega)\log(\frac{H}{\delta})$. Substituting eqs.(\ref{H1Miao},\ref{H2Miao}) into the entropy functional (\ref{HEEMiao3d3}), we obtain
\begin{eqnarray}\label{aHCG}
a(h_0)&=&(-\frac{f_0}{6}+4  c^{(2)}_1)a_E(h_0)+\sum_{m=2}^{\infty}\lambda_m a_{K,m}(h_0)+O(\lambda_m)^2
\end{eqnarray}
where $a_E\sim 1/h_0^2$ is given by eq.(\ref{aEin}), and  $a_m(h_0)$ can be derived as
\begin{eqnarray}\label{amHCG}
a_{K,m}&=&-\frac{1}{4G}\int_{0}^{\infty}dy\frac{4  \left(\left(4 m^2+1\right) \left(y^2+1\right)-\frac{m \left(\left(y^2+1\right)^{2 m}-1\right)}{y^2}-m \left(3 y^2+2\right)\right)}{h_0^{2 m}\left(y^2+1\right)^{2 m+1}}+O(\frac{1}{h_0^{2m+2}})\nonumber\\
&=&-\frac{1}{4G}\frac{\bar{a}_m\pi}{h_0^{2m}}+O(\frac{1}{h_0^{2m+2}})
\end{eqnarray}
 in the smooth limit $h_0\to \infty$. Here $\bar{a}_m$ are some finite numbers:
 \begin{eqnarray}\label{baram}
\bar{a}_1=1,\ \bar{a}_2=17/8,\ \bar{a}_3=453/128,\ \bar{a}_4=5189/1024,\ \bar{a}_5=218285/32768,...
 \end{eqnarray}
Similar to $\Omega_m$ (\ref{angleHCGm}), $a_{K,m}\sim 1/h_0^{2m} (m\ge 2)$ are subleading terms. Thus, the higher extrinsic curvature terms can be ignored if we focus on the leading terms of $\Omega(h_0)$ and $a(h_0)$. It should be mentioned that the equivalence between $\sqrt{\gamma} Tr K^2$ and $-2\sqrt{\gamma} $ implies $\bar{a}_1=1$, which is consistent with eq.(\ref{baram}). As another check of our results, we have calculated $\Omega(h_0)$ and $a(h_0)$ numerically and got perfect agreements with the exact results eqs.(\ref{angleHCG}-\ref{baram}) for large $h_0$. 

Now we are ready to derive $\sigma$ for the general higher curvature gravity. Recall that  $\sigma$ is defined by $a(\Omega)$ in the smooth limit:
 \begin{eqnarray}\label{recallsigma}
 \sigma=\lim\limits_{\Omega\to\pi}\frac{a(\Omega)}{(\pi-\Omega)^2}
 \end{eqnarray}
From eqs.(\ref{angleHCG},\ref{angleHCGm},\ref{aHCG},\ref{amHCG}), we get
 \begin{eqnarray}\label{adifference}
 a(\Omega)=(-\frac{f_0}{6}+4  c^{(2)}_1)a_E(\Omega)+\lambda_m \ O(\pi-\Omega)^4+O(\lambda_m)^2
 \end{eqnarray}
 Thus we have 
 \begin{eqnarray}\label{proofsigma}
 \sigma=(-\frac{f_0}{6}+4  c^{(2)}_1)\sigma_E
 \end{eqnarray}
 at least up to order $O(\lambda_m)$. Recall that the central charge is given by
 \begin{eqnarray}\label{proofCT}
 C_T=(-\frac{f_0}{6}+4  c^{(2)}_1)C_{E,T}
 \end{eqnarray}
We obtain
\begin{eqnarray}\label{proofconjectHCG}
\frac{\sigma}{C_T}=\frac{\sigma_E}{C_{T,E}}=\frac{\pi^2}{24}.
\end{eqnarray}
Now we finish the proof of the conjecture (\ref{conjecture}).

One may wonder what happens if we take into account the higher orders terms $O( \lambda_m)^n$ in our perturbative approach. It turns out that these higher order terms decrease quickly as 
\begin{eqnarray}\label{powerlawOmega}
\Omega-\alpha\Omega_E &\sim& \sum_{m,n} \frac{\lambda_m^n }{h_0^{(2m-2)n+1}}\\
a(h_0)-\alpha a_E(h_0)&\sim& \sum_{m,n} \frac{\lambda_m^n }{h_0^{(2m-2)n+2}}\label{powerlawa}
\end{eqnarray}
where $\alpha=(-\frac{f_0}{6}+4  c^{(2)}_1)$. 
Thus, the higher order terms $O( \lambda_m)^n$ are less important than the first order terms $O(\lambda_m)$, and they do not change $\sigma$ either. A quick 'derivation' of the power law eqs.(\ref{powerlawOmega},\ref{powerlawa}) is as follows. Note that 
\begin{eqnarray}\label{dh}
&&dh\sim dy\nonumber\\
&&h'\sim h_0\bigg(c_0(y) +\sum_{m=2}\frac{c_m(y)+\lambda_m}{h_0^{2m-2}}\bigg)\nonumber\\
&&L\sim\frac{1}{h_0}
\end{eqnarray}
where we have subtracted the divergent parts in $L$. To see why $L\sim\frac{1}{h_0}$, it is helpful to note that $a(h_0)\sim L/h'\sim 1/h_0^2$.
Then we get
\begin{eqnarray}\label{powerlawOmega1}
\Omega-\alpha\Omega_E &\sim& \int dy (\frac{1}{h'}-\frac{\alpha}{h_E'})\sim \sum_{m,n} \frac{\lambda_m^n }{h_0^{(2m-2)n+1}} \\
a(h_0)-\alpha a_E(h_0)&\sim& \int dy (\frac{L}{h'}-\frac{\alpha L_E}{h_E'}) \sim \sum_{m,n} \frac{\lambda_m^n }{h_0^{(2m-2)n+2}}\label{powerlawa1}
\end{eqnarray}
In the appendix, we calculate the $O(\lambda_m)^2$ terms exactly and find that they indeed obey the power law eqs.(\ref{powerlawOmega},\ref{powerlawa}). We have also checked some  $O(\lambda_m)^3$ terms, which satisfy the power law eqs.(\ref{powerlawOmega},\ref{powerlawa}) too. Thus, similar to the smooth case, the higher extrinsic curvature terms $(K^{2m})$ with $m\ge 2$ can be ignored in the smooth limit. $\sigma$ is irrelevant with these terms to arbitrary order $O(\lambda^n_m)$.

To sum up, we give a holographic proof of the conjecture (\ref{conjecture}) for the CFTs dual to the general higher curvature gravity. We work in the perturbative framework with $\lambda_m\ll 1$ in order to suppress the massive modes and to have a well-defined extremal entropy surface for higher curvature gravity.

\subsection{Discussions}

In this subsection, we discuss some interesting questions raised by \cite{Myers0,Myers1}. Firstly, we show that the lower bound $a_E(\Omega)/C_T$ associated to Einstein gravity can be violated by higher curvature gravity. Secondly, we find that, in contrast to $\sigma/C_T$, $\kappa/C_T$ is not universal. In general, $\kappa$ depends on infinite parameters of the higher curvature gravity. Thus, it is not a good candidate for the central charge. Let us discuss the above two problems one by one below.

For the first problem, we set $\lambda_m=0$ with $m\ge 3$ for simplicity. This means that we focus on the higher curvature gravity including at most the cubic curvature terms. Using eqs.(\ref{angleHCG}-\ref{baram}), we obtain
\begin{eqnarray}\label{lowerbound}
\frac{a(\Omega)}{C_T}-\frac{a_E(\Omega)}{C_{T,E}}=-\frac{5}{96}\bar{\lambda}_2 (\pi-\Omega)^4+O(\pi-\Omega)^6
\end{eqnarray}
Now it is clear that $\frac{a(\Omega)}{C_T}$ with positive $\bar{\lambda}_2=\lambda_2/\alpha_3$ is smaller than $\frac{a_E(\Omega)}{C_{T,E}}$ near $\Omega\sim\pi$  \footnote{We notice that $\bar{\lambda}_2$ is proportional to $t_4$, which is the parameter of three point functions for the stress tensor. We have a constraint $-4\le t_4\le 4$ from the positivity of energy \cite{Myers4}. Thus, $\bar{\lambda}_2\sim t_4$ can indeed be positive. }. Thus, the lower bound $a_E(\Omega)/C_T$ associated to Einstein gravity can indeed be violated by higher curvature gravity.

Now let us go on to discuss the second problem. Recall that $\kappa$ is defined in the small angle limit of $a(\Omega)$:
\begin{eqnarray}\label{kappa0}
\kappa=\lim\limits_{\Omega\to 0} a(\Omega)\ \Omega
\end{eqnarray}
Following the approach of last subsection, we can express the opening angle $\Omega$ as eq.(\ref{angleHCG})
\begin{eqnarray}\label{angleHCGkappa}
\Omega=\Omega_E(h_0)+\sum_{m=2}^{\infty}\bar{\lambda}_{m} \Omega_m(h_0)+O(\bar{\lambda}_{m})^2
\end{eqnarray}
with $\Omega_m(h_0\to 0)$ given by
\begin{eqnarray}\label{angleHCGmR0}
\Omega_m&=&\int_{0}^{\infty}dy\frac{2 \left(2 (1-2 m) r_0^2+2 r_0^{4 m-2} \left((2 m-1) r_0^4+2 m (4 m-3) r_0^2 y^2+m (4 m-3) y^4\right) \left(r_0^2+y^2\right)^{-2 m}\right)}{y^2 \left(2 r_0^2+y^2\right)^{3/2} (r_0^2+y^2)^{-1/2}}\nonumber\\
&&+O(\frac{1}{r_0^3})\nonumber\\
&=& \hat{\omega}_m\pi ^{3/2} h_0 + O(h_0^3).
\end{eqnarray}
 Here we have substituted $r_0=1/h_0$ and $\hat{\omega}_m$ are some finite numbers given by
\begin{eqnarray}\label{omegamR0}
\hat{\omega}_1=0,\ \hat{\omega}_2=\frac{128 \sqrt{2} }{15 \Gamma \left(-\frac{3}{4}\right)^2},\ \hat{\omega}_3=\frac{2176 \sqrt{2} }{135 \Gamma \left(-\frac{3}{4}\right)^2},\ \hat{\omega}_4=\frac{644 \sqrt{2} }{195 \Gamma \left(\frac{1}{4}\right) \Gamma \left(\frac{5}{4}\right)},\ \hat{\omega}_5=\frac{956 \sqrt{2}}{221 \Gamma \left(\frac{1}{4}\right) \Gamma \left(\frac{5}{4}\right)},...
\end{eqnarray}
From eqs.(\ref{smallangle0},\ref{angleHCGkappa},\ref{angleHCGmR0}), we obtain
\begin{eqnarray}\label{angleHCGkappa1}
\Omega=\bigg(\frac{2\sqrt{\pi}\Gamma(\frac{3}{4})}{\Gamma(\frac{1}{4})}+\sum_{m=2}^{\infty}\bar{\lambda}_{m} \hat{\omega}_m\pi ^{3/2}+O(\bar{\lambda}_{m})^2\bigg)h_0+O(h_0^3)
\end{eqnarray}
Remarkably, all the higher curvature terms contribute to the leading term of $\Omega$ in the singular limit.

Similarly, we can derive $a(h_0)$ as eq.(\ref{aHCG})

\begin{eqnarray}\label{aHCGkappa}
a(h_0)&=&(-\frac{f_0}{6}+4 c^2_1)\bigg(a_E(h_0)+\sum_{m=2}^{\infty}\bar{\lambda}_m a_{K,m}(h_0)+O(\bar{\lambda}_m)^2\bigg)
\end{eqnarray}
with $a_{K,m}(h_0\to 0)$ given by
\begin{eqnarray}\label{amHCGkappa}
a_{K,m}&=&-\frac{1}{4G}\int_{0}^{\infty}dy\frac{4 \left(r_0^{4 m} \left((2 m-1) r_0^4+2 (m (4 m-3)+1) r_0^2 y^2+(m (4 m-3)+1) y^4\right)-(2 m-1) r_0^4 \left(r_0^2+y^2\right)^{2 m}\right)}{y^2 \left(2 r_0^2+y^2\right)^{3/2} \left(r_0^2+y^2\right)^{2 m-\frac{1}{2}}}\nonumber\\
&&+O(\frac{1}{r_0})\nonumber\\
&=&-\frac{1}{4G}\frac{\hat{a}_m\pi ^{3/2}}{h_0}+O(h_0)
\end{eqnarray}
where $r_0=1/h_0$ and $\hat{a}_m$ are some finite numbers:
\begin{eqnarray}\label{hatam}
\hat{a}_1=\frac{4  \Gamma \left(\frac{3}{4}\right)}{\pi\Gamma \left(\frac{1}{4}\right)},\ \hat{a}_2=-\frac{12 \sqrt{2} }{5 \Gamma \left(-\frac{3}{4}\right) \Gamma \left(\frac{5}{4}\right)},\ \hat{a}_3=\frac{2624 \sqrt{2} }{135 \Gamma \left(-\frac{3}{4}\right)^2},\ \hat{a}_4=\frac{2884 \sqrt{2} }{195 \Gamma \left(\frac{1}{4}\right)^2},\ \hat{a}_5=\frac{66112 \sqrt{2} }{1989 \Gamma \left(-\frac{3}{4}\right)^2},...
\end{eqnarray}
From eqs.(\ref{smalla0},\ref{angleHCGkappa1},\ref{aHCGkappa},\ref{amHCGkappa}), we obtain
\begin{eqnarray}\label{a}
a(\Omega)&=&(-\frac{f_0}{6}+4  c^{(2)}_1)\bigg(\frac{\Gamma \left(\frac{3}{4}\right)^4}{2 \pi  G}+\sum_{m=2}^{\infty}\bar{\lambda}_m\frac{\pi ^2   \Gamma \left(-\frac{1}{4}\right) (\hat{a}_m-\hat{\omega}_m)}{8 G \Gamma \left(\frac{1}{4}\right)}+O(\bar{\lambda}_m)^2\bigg) \frac{1}{\Omega}+O(\Omega)
\end{eqnarray}
Thus, $\kappa$ depends on all the parameters of the higher curvature gravity. As a result, $\kappa/C_T$ is not a universal ratio:
\begin{eqnarray}\label{kappaCT}
\frac{\kappa}{C_T}=\frac{1}{6} \pi ^2 \Gamma \left(\frac{3}{4}\right)^4+\sum_{m=2}^{\infty}\bar{\lambda}_m\frac{\pi ^5 \Gamma \left(-\frac{1}{4}\right) (\hat{a}_m-\hat{\omega}_m)}{24 \Gamma \left(\frac{1}{4}\right)}+O(\bar{\lambda}_m)^2.
\end{eqnarray}
The holographic models studied by \cite{Myers0,Myers1} imply that $\kappa/C_T$ seems to be a universal ratio. However, as we have shown here, this is not the case for general holographic models. Our results explain the field theoretical mismatch of the ratio $\kappa/C_T$ between free scalar (4.17945) and free fermion (3.8005) \cite{Myers1}. Eq.(\ref{kappaCT}) shows that $\frac{\kappa}{C_T}$ crucially depends on the parameters of the holographic models, or equivalently, the details of CFTs. Thus, there is no reason to expect  $\frac{\kappa}{C_T}$ to be the same for scalars and fermions.

\section{New conjectures for CFTs in higher dimensions}

In this section, we investigate the universal contributions to EE from high-dimensional cones. On the gravity side, we focus on the following AdS metric
\begin{eqnarray}\label{Hcones1}
	ds^2=\frac{dz^2+dt^2_E+d\rho^2+\rho^2(d\theta^2+\sin^2\theta d\Omega^2_{d-3})}{z^2}
\end{eqnarray}
where $\theta\in [0, \Omega]$ with $\Omega \le \pi$ and $d\Omega^2_{d-3}$ is the metric of unit $(d-3)$-sphere. According to \cite{Myers3,Safdi}, the universal terms of EE are logarithmic terms $-a_d(\Omega)\log(H/\delta)$ and squared logarithmic terms $-a_d(\Omega)\log^2(H/\delta)$ in odd dimensions and even dimensions, respectively. Similar to the 3d CFTs, we have $a_d(\Omega)=a_d(\pi-\Omega)$ and the following asymptotic behaviors
\begin{eqnarray}\label{Hconesbe1}
	a_d(\Omega\to \pi/2)\simeq \sigma_d (\pi/2-\Omega)^2,\ \ \ a_d(\Omega\to 0)\simeq \kappa_d/\Omega
\end{eqnarray}
By studying the holographic models, we find that
\begin{eqnarray}\label{conjecture21}
\frac{\sigma_d}{C_T}=\frac{\sigma_{d,E}}{C_{T,E}}
\end{eqnarray}
is a universal ratio. Here $C_T$ are the central charges defined in the two point function eq.(\ref{twopoint}) and $E$ denotes Einstein gravity.

\subsection{CFTs in even dimensions}

\subsubsection{4d CFTs}

In this subsection, we study the universal terms of EE from sharp corners for 4d CFTs. For simplicity, we firstly consider gravity theories with at most squared curvatures and then generalize our discussions to general higher curvature gravity. 

From eq.(\ref{HEEMiao}), we get the entropy for curvature-squared gravity as
\begin{eqnarray}\label{HEE4dR2}
S=\frac{1}{4 G}\int d^{d-1}y \sqrt{\gamma}\big{[} -\frac{f_0}{2d}- c^{(2)}_2\frac{1}{2}(tr K)^2-2 c^{(2)}_1 tr K^2 \big{]}
\end{eqnarray} 
As argued in \cite{Myers0,Myers1}, we can drop $(tr K)^2$ near the minimal surface. Take into account the scaling symmetry of $AdS$, we can parameterize the extremal entropy surface as $z= \rho\  h(\theta)$. Now the entropy functional becomes
\begin{eqnarray}\label{HEE4dR21}
S=\frac{\Omega_{d-3}}{4 G}\int_{\delta/h_0}^{H}\frac{d \rho}{\rho}\int_{h_0}^{\delta/\rho}dh \frac{\sin^{d-3}(\theta)}{h'h^{d-1}}\sqrt{1+h^2+(h')^2}\bigg(-\frac{f_0}{2d}-2 c^{(2)}_1 \big(k^2_++k^2_-+(d-3)k^2_0\big) \bigg)
\end{eqnarray} 
where $k_{\pm}, k_0$ are the eigenvalues of the extrinsic curvature $K^i_j$:
\begin{eqnarray}\label{eigenvalues}
k_+&=&\frac{1}{\sqrt{h^2+\left(h'\right)^2+1}},\\
k_0&=&\frac{h^2+h h' \cot (\theta)+1}{\sqrt{h^2+\left(h'\right)^2+1}},\\
k_-&=&\frac{h^4+h h''+2 h^2+\left(h'\right)^2+h^3 h''+1}{\left(h^2+\left(h'\right)^2+1\right)^{3/2}}
\end{eqnarray} 

Now let us consider the case $d=4$. Firstly, we derive E.O.M of $h(\theta)$ from the entropy functional (\ref{HEE4dR21}). Then we change the variable $y=\sin(\theta)=y(h)$. Finally, we solve E.O.M of $y(h)$ perturbatively. After some tedious calculations, we obtain 
\begin{eqnarray}\label{4dsolution}
y=\sin (\Omega )-\frac{1}{4} \cos (\Omega ) \cot (\Omega )h^2+y_0 h^4-\frac{1}{64}  (\cos (2 \Omega )-3) \cot ^2(\Omega ) \csc (\Omega ) h^4 \log (h)+O(h^6)
\end{eqnarray} 
where $y_0$ is a constant that can be fixed by using the fact that $y(h)$ has an extrema at $h=h_0$. Remarkably, the solution (\ref{4dsolution}) is independent of the parameter $c^{(2)}_1$ up to $O(h^4)$. This means that the minimal surface is a good approximation for the extremal entropy surface near the boundary of $AdS$. Using the solution (\ref{4dsolution}) together with $h'=\sqrt{1-y^2}/y'(h)$ and $h''=-(yy'^2+(1-y^2)y'')/y'^3$, we find the integrand of eq.(\ref{HEE4dR21}) behaves as
\begin{eqnarray}\label{4ddivergentterm}
\frac{\sin^{d-3}(\theta)}{h'h^{d-1}}\sqrt{1+h^2+(h')^2}\bigg(-\frac{f_0}{2d}-2 c^{(2)}_1 tr K^2 \bigg)=\frac{f_0}{8}\frac{\sin(\Omega)}{h^3}+(-\frac{f_0}{8}+8 c^{(2)}_1)\frac{\cos (\Omega ) \cot (\Omega )}{8h}+O(h)
\end{eqnarray} 
According to \cite{Myers3}, the universal squared logarithmic term can only come from the ($1/h$) term in the integrand (\ref{4ddivergentterm}). Substituting eq.(\ref{4ddivergentterm}) into entropy functional (\ref{HEE4dR21}), we get the universal term of EE as 
\begin{eqnarray}\label{4dloglogR2}
-a_4(\Omega)\log^2(H/\delta)=-\frac{\pi}{32G}(-\frac{f_0}{8}+8 c^{(2)}_1)\cos (\Omega ) \cot (\Omega )\log^2(H/\delta)
\end{eqnarray}

Now let us generalize our above discussions to general higher curvature gravity. It turns out that the general holographic models give the same result as eq.(\ref{4dloglogR2}). The reasons are as follows. Near the boundary $h\to 0$, the asymptotic solution $y(h)$ takes the form 
\begin{eqnarray}\label{2ndsolution}
y=\sin (\Omega )+c_1 h^2+...+ h^{d}(c_{d/2}+b \log(h))+...
\end{eqnarray} 
Thus, we have $h'=\sqrt{1-y^2}/y'(h)\sim 1/h$ and $h''=-(yy'^2+(1-y^2)y'')/y'^3\sim 1/h$. Substituting $h'\sim 1/h$ and  $h''\sim 1/h$ into the eigenvalues of the extrinsic curvature (\ref{eigenvalues}), we find that 
\begin{eqnarray}\label{Kpowers}
(K^{2m})\sim h^{2m}
\end{eqnarray} 
Thus, the entropy functional for higher extrinsic curvature terms $(K^{2m})$ behaves 
\begin{eqnarray}\label{LKpowers}
\int\frac{d \rho}{\rho}\int^{\delta/\rho}dh \frac{\sin^{d-3}(\theta)}{h'h^{d-1}}\sqrt{1+h^2+(h')^2}\ (K)^{2m}\sim \int\frac{d \rho}{\rho}\int^{\delta/\rho}dh \big( h^{2m-d+1}+...\big)
\end{eqnarray}
where '...' denotes higher order terms. 
Now it is clear only the terms $(K^{2m})$ with $m\le (d-2)/2$ contribute to the squared logarithmic terms. That is because only the $1/h$ terms in the integrand are related to the universal term of EE \cite{Myers3}. For $d=4$, we get $m\le 1$. Thus, besides $Tr K^2$, there is no need to consider other higher extrinsic curvature terms for 4d CFTs.

Using eq.(\ref{4dloglogR2}) together with $C_T=(-\frac{f_0}{8}+8 c^{(2)}_1)C_{T, E}$ for $d=4$, we find that 
\begin{eqnarray}\label{4dConjecture}
\frac{a_4(\Omega)}{C_T}=\frac{a_{4,E}(\Omega)}{C_{T,E}}
\end{eqnarray} 
is a universal ratio for all the CFTs dual to higher curvature gravity. 
Note that we have not only $\sigma_4/C_T$ but also $a_4(\Omega)/C_T$ universal for 4d CFTs.

\subsubsection{6d CFTs}

Now let us study the universal terms of EE from sharp corners for 6d CFTs. As discussed below eq.(\ref{LKpowers}), to derive $a_6(\Omega)$, we only need to consider three extrinsic curvature terms, i.e., $Tr K^2$, $Tr K^4$ and $(tr K^2)^2$ in the entropy functional (\ref{HEEMiao}). Note that, similar to $(tr K)^2$, $(tr K)(tr K^3)$ and $(tr K)^4$ are less important near the boundary of AdS, where the extremal entropy surface becomes the minimal surface approximately. 

Let us firstly consider $Tr K^2$. The corresponding entropy functional is given by eq.(\ref{HEE4dR21}). Following the approach of last subsection, we obtain the asymptotic solution
\begin{eqnarray}\label{6dsolution}
y&=&\sin (\Omega )-\frac{3}{8}  \cos (\Omega ) \cot (\Omega )h^2-\frac{3 \cot ^2(\Omega ) \csc (\Omega ) ((104 \lambda +11) \cos (2 \Omega )+168 \lambda +19)}{1024 (8 \lambda +1)}h^4 \nonumber\\
&&+y_0 h^6-\frac{3  \cos (\Omega ) \cot (\Omega ) \left((112 \lambda -15) \csc ^4(\Omega )+2 (48 \lambda -7) \csc ^2(\Omega )+48 \lambda -3\right)}{4096 (8 \lambda -1)}h^6 \log (h)\nonumber\\
&&+O(h^8)
\end{eqnarray}
For simplicity, we have set $-\frac{f_0}{2d}=1$ and $\lambda=-2c^{(2)}_1$ in the above equation. 
Substituting the solution (\ref{6dsolution}) into the entropy functional (\ref{HEE4dR21}) and picking the $1/h$ terms in the integrand, we derive 
\begin{eqnarray}\label{6dloglogR2}
a_6(\Omega)&=&\frac{\Omega_3}{8G}\frac{9 \cos (\Omega ) \cot (\Omega ) ((1-16 \lambda ) \cos (2 \Omega )+240 \lambda -31)}{4096}\\
\sigma_6&=&-\frac{\Omega_3}{8G}\frac{9}{128} (1-8 \lambda)\label{6dsigmaR2}
\end{eqnarray}

Let us go on to discuss the effects from $Tr K^4$ and $(tr K^2)^2$. From the experience of 3d and 4d CFTs, it is expected that $Tr K^4$ and $(tr K^2)^2$ do not change $\sigma_6$. Instead, they only modify the subleading terms of $a_6(\Omega)$ in the smooth limit $\Omega\to \pi/2$. As we shall show below, this is indeed the case. Adding $\lambda_{2(1)}tr K^4+\lambda_{2(2)}(tr K^2)^2$
to the  the entropy functional (\ref{HEE4dR2}) and following the above approach, we find that $tr K^4$ and $(tr K^2)^2$ only modify $a_6(\Omega)$ at the subleading order $O(\Omega -\frac{\pi }{2})^4$
\begin{eqnarray}\label{6daK4}
\delta a_6(\Omega )=\frac{\Omega_3}{8G} (\Omega -\frac{\pi }{2})^4\bigg(-\frac{21}{64}\lambda_{2(1)}-\frac{9}{16}\lambda_{2(2)}\bigg)+ O(\Omega -\frac{\pi }{2})^6
\end{eqnarray}
Thus, the coefficient of $(\Omega -\frac{\pi }{2})^2$, i.e., $\sigma_6$ (\ref{6dsigmaR2}), remains the same. Now from eq.(\ref{6dsigmaR2}) and $C_T=(1-8 \lambda)C_{T,E}$, it is clear that 
\begin{eqnarray}\label{6dConjecture}
\frac{\sigma_6}{C_T}=\frac{\sigma_{6,E}}{C_{T,E}}
\end{eqnarray} 
is indeed a universal ratio for the CFTs dual to general higher curvature gravity. 

\subsubsection{2n-dimensional CFTs}

Now let us investigate the universal term of EE from sharp corners for general even-dimensional CFTs. The higher the dimension is, the more terms we need to consider in order to derive $a_d(\Omega)$. In general, we need to study all the extrinsic curvature terms  $(K^{2m})$ with $m\le (d-2)/2$ for $a_d(\Omega)$. The experiences of 4d and 6d CFTs imply that only $tr K^2$ contribute to $\sigma_d$ and the other higher extrinsic curvature terms only modify the subleading terms of $a_d(\Omega)$. For simplicity, we only consider the curvature-squared gravity in this sub-section. 

Using the entropy functional (\ref{HEE4dR21}) and following the approach of sect. 5.1.1, we obtain
\begin{eqnarray}
\sigma_d&=& \frac{\Omega_{d-3}}{8G}  \big(-\frac{f_0}{2d}+(4d-8) c^{(2)}_1\big)\ \beta_d \label{2ndsigmaR2}
\end{eqnarray}
where $\Omega_{d-3}=\frac{2\pi^{(d-2)/2}}{\Gamma({(d-2)}/2)}$ is the volume of the unit $(d-3)$-sphere, $\beta_d$ are some numbers given by
\begin{eqnarray}\label{betad}
\beta_4=\frac{1}{8},\ \ \beta_6=-\frac{9}{128},\ \  \beta_8=\frac{25}{512},\ \  \beta_{10}=-\frac{1225}{32768},\  \  \beta_{12}=\frac{3969}{131072}, \ ...
\end{eqnarray}
We notice that $\beta_d$ are the expansion coefficients of complete elliptic integral of the first kind:
\begin{eqnarray}\label{ellipticintegral}
&&K(-x)=\frac{\pi }{2}-\pi\sum_{n=2}^{\infty} \beta_{2n} \ x^{n-1} \\
&& \beta_{d}=(-1)^{\frac{d}{2}}\frac{2^{3-2d}\Gamma[d-1]^2}{\Gamma[\frac{d}{2}]^2}
\end{eqnarray}

Comparing $\sigma_d$ (\ref{2ndsigmaR2}) with $C_T=(-\frac{f_0}{2d}+(4d-8) c^{(2)}_1)C_{T,E}$, we find that
\begin{eqnarray}\label{2ndConjecture}
\frac{\sigma_d}{C_T}=\frac{\Omega_{d-3} }{8G}\frac{\beta_d}{C_{T,E}}=(-1)^{\frac{d}{2}}\frac{(d-1)(d-2)\pi^{d-1}\Gamma[\frac{d-1}{2}]^2}{2 \Gamma[\frac{d}{2}]^2\Gamma[d+2]},\ \ \text{(d even)}
\end{eqnarray}
is a universal ratio for even-dimensional CFTs \footnote{Note that the definition of $\sigma_d$ of this paper differs from the one of \cite{Myers5} by a factor $4(-1)^{\frac{d}{2}}$. Thus, eq.(\ref{2ndConjecture}) agrees with the results of \cite{Myers5}.}. Although we only checked eq.(\ref{2ndConjecture}) by studying the curvature-squared gravity, we expect that it is a universal law for the CFTs dual to general higher curvature gravity in space-time (\ref{Hcones1}). It should be mentioned that $a_d(\Omega)$ is not conformally invariant for even CFTs \cite{Myers3}. To derive eq.(\ref{2ndConjecture}), we assume that the boundary metric is 
\begin{eqnarray}\label{CFTspacetime1}
ds^2=dt^2_E+d\rho^2+\rho^2(d\theta^2+\sin^2\theta d\Omega^2_{d-3})
\end{eqnarray}
which is dual to the bulk metric (\ref{Hcones1}). By a singular Weyl transformation, we can obtain a new boundary metric \cite{Myers3}
\begin{eqnarray}\label{CFTspacetime2}
ds^2=dY^2+d\xi^2+\sin^2\xi(d\theta^2+\sin^2\theta d\Omega^2_{d-3})
\end{eqnarray}
which is dual to the bulk metric
\begin{eqnarray}\label{bulkmetric2}
ds^2=\frac{1}{1+R^2}dR^2+(1+R^2)dY^2+R^2[d\xi^2+\sin^2\xi(d\theta^2+\sin^2\theta d\Omega^2_{d-3})]
\end{eqnarray}
It turns out that $a_d(\Omega)$ derived from (\ref{Hcones1}) and (\ref{bulkmetric2}) differ by a factor 2. This mismatch can be regarded as an anomaly from the singular conformal transformation \cite{Myers3}. Thus, by saying $\sigma_d/C_T$ (\ref{2ndConjecture}) is universal for even-dimensional CFTs, we mean the case when all the CFTs live in the same boundary space-time. 

\subsection{CFTs in odd dimensions}

Now let us study the universal term of EE from sharp corners for CFTs in odd dimensions. In contrast to 3d and even-dimensional cases, it is difficult to derive the exact formula of $\sigma_d$ for odd-dimensional CFTs. We leave the discussions of odd-dimensional CFTs to future work. Below we proceed as far as we can. For simplicity, we take 5d CFTs as an example. 

Let us start with the entropy functional (\ref{HEEMiao}) with $d=5$, $-f_0/10=1$ and $\lambda=-2c^{(2)}_1$.
\begin{eqnarray}\label{HEE5dR21}
S=\frac{\Omega_{2}}{4 G}\int_{\delta/h_0}^{H}\frac{d \rho}{\rho}\int_{h_0}^{\delta/\rho}dh \frac{\sin^{2}(\theta)}{h'h^{4}}\sqrt{1+h^2+(h')^2}\bigg(1+\lambda tr K^2 +\sum_{m=2}^{\infty}\lambda_m(K^{2m})\bigg)
\end{eqnarray} 
where $tr K^2=\big(k^2_++k^2_-+2k^2_0\big)$ with $k$ given by eq.(\ref{eigenvalues}), and $(K^{2m})$ denote all the possible higher extrinsic curvature terms. Following the approach of sect.5.1.1, we can derive the asymptotic solution 
\begin{eqnarray}\label{5dsolution}
y=\sin (\Omega )-\frac{1}{3} h^2 \cos (\Omega ) \cot (\Omega )-\frac{h^4 \left(\cot ^2(\Omega ) \csc (\Omega ) ((5 \lambda +1) \cos (2 \Omega )+11 \lambda +4)\right)}{54 (2 \lambda +1)}+O(h^6)
\end{eqnarray} 
Remarkably, the higher extrinsic curvature terms $(K^{2m})$ with $m\ge 2$ do not affect the asymptotic solution up to order $O(h^4)$. That is because  $(K^{2m})\sim h^{2m}$ are subleading terms near the boundary. This is a sign that these higher extrinsic curvature terms are irrelevant to $\sigma_5$.

Substituting the above solution into the integrand of the entropy functional (\ref{HEE5dR21}), we find
\begin{eqnarray}\label{5ddivergentterm}
\frac{\sin^{2}(\theta)}{h'h^{4}}\sqrt{1+h^2+(h')^2}\bigg(1+\lambda tr K^2+\sum_{m=2}^{\infty}\lambda_m(K^{2m}) \bigg)=-\frac{\sin(\Omega)}{h^4}-\frac{2 (3 \lambda -2) \cos ^2(\Omega )}{9 h^2}+O(1)
\end{eqnarray} 
To derive the universal term of EE $-a_5(\Omega)\log(H/\delta)$, we need to extract the finite part of the above integrand. We obtain
\begin{eqnarray}\label{5da}
a_5&=&\frac{\pi}{G}(\frac{\sin(\Omega)}{3h_0^3}+\frac{2 (3 \lambda -2) \cos ^2(\Omega )}{9 h_0})\nonumber\\
&+&\frac{\pi}{G}\int_{0}^{h_0}dh\bigg(\frac{\sin^{2}(\theta)}{h'h^{4}}\sqrt{1+h^2+(h')^2}\big(1+\lambda tr K^2+\sum_{m=2}^{\infty}\lambda_m(K^{2m}) \big)+\frac{\sin(\Omega)}{h^4}+\frac{2 (3 \lambda -2) \cos ^2(\Omega )}{9 h^2}\bigg)\nonumber\\
\end{eqnarray} 
Unlike the 3d case, it is difficult to derive an analytical solution of $h(\theta)$ and thus $a_5$ due to the appearance of $\sin(\theta)$ in the entropy functional (\ref{HEE5dR21}). We leave the careful numerical study of this problem to future work \cite{Li}.

\section{Conclusions}

 By applying the general higher curvature gravity, we give a holographic proof of the conjecture \cite{Myers0} for 3d CFTs. We find that, similar to the smooth case, the cubic and higher terms in the expansions of the action (\ref{GHCaction0}) around the 'background-curvature' are less important, i.e., they do not change either $\sigma$ or $C_T$. Besides, we have clarified some interesting problems. Firstly, we find that, unlike $\sigma/C_T$, $\kappa/C_T$ is not a universal ratio. On the contrary, it crucially depends on the details of the CFTs. Secondly, we find that the lower bound $a_E(\Omega)/C_T$ associated to Einstein gravity can be violated by higher curvature gravity. Last but not least, we find that there are similar universal laws in the smooth limit for CFTs in higher dimensions. We give a holographic proof of the universal laws for 4d and 6d CFTs which are dual to the general higher curvature gravity and check the higher even-dimensional cases by studying curvature-squared gravity. As for the odd-dimensional CFTs ($d>3$), it is difficult to derive the analytical results. However there are hints that the higher extrinsic curvature terms do not affect $\sigma$. Therefore it is expected that, similar to the 3d CFTs, $\sigma_d/C_{T,d}$ are also universal ratios for the CFTs in higher odd-dimensional space-times. We leave the careful numerical study of this problem to future work \cite{Li}. Based on the holographic results, we can trust these new conjectures in higher dimensions at least for strongly coupled CFTs. It is interesting to test whether these universal laws are obeyed by weakly coupled CFTs. It is also interesting to find a field theoretical proof of these universal ratios. Finally, we want to mention that, for simplicity, we focus on the CFTs dual to the general higher curvature gravity $f(R_{ijkl})$ in this paper. It is expected that our discussions can be naturally generalized to the cases of most general higher derivative gravity $f(R_{ijkl}, \nabla_mR_{ijkl},...)$. Now work is in progress in this direction.

\section*{Acknowledgements}

R. X. Miao is supported by Sino-German (CSC-DAAD) Postdoc Scholarship Program. R. X. Miao thank S. Theisen for helpful discussions.

\appendix

\section{Some formula}
\begin{eqnarray}\label{FMiao}
&& F_m=\frac{h^2+2 m \left(h'\right)^2+1}{h^2 \left(h^2+\left(h'\right)^2+1\right)^{m+\frac{1}{2}}}\nonumber\\
&&+\frac{\left(h^2+1\right)^2 \left((h-6 h m) h''+h^2+1\right)+2 \left(2 h^2+3\right) m \left(h'\right)^4+\left(h^2+1\right) \left(2 \left(5 h^2+3\right) m+1\right) \left(h'\right)^2}{h^2 \left(h^2+\left(h'\right)^2+1\right)^{3 m+\frac{1}{2}} \left(\left(h^2+1\right) \left(h h''+h^2+1\right)+\left(h'\right)^2\right)^{1-2 m}}\nonumber\\
&&+\frac{2 m \left(\left(h^2+1\right) \left(h h''+h^2+1\right)+\left(h'\right)^2\right)^{2 m-2}}{h^2 \left(h^2+\left(h'\right)^2+1\right)^{\frac{1}{2}-3 m}}\times \bigg(2 h \left(h^2+1\right)^3 h'' \left(h h''+h^2+1\right)-\left(h^2+3\right) \left(h'\right)^6\nonumber\\
&&\ \ \ \ +\left(h^2+1\right) \left(h'\right)^4 \left(h^4 (8 m-5)+h \left(h^2 (6 m-4)-3\right) h''+2 h^2 (m-4)-6\right)\nonumber\\
&&\ \ \ \ +\left(h^2+1\right)^2 \left(h'\right)^2 \left(\left(h^2+1\right) \left(2 h^2 (m-2)-3\right)-h h'' \left(3 h (2 m-1) h''+6 h^2 m+1\right)\right)\nonumber\\
&&\ \ \ \ +\left(h^3+h\right)^2 h^{(3)} (2 m-1) \left(h'\right)^3+h^2 h^{(3)} \left(h^2+1\right)^3 (2 m-1) h'\bigg)
\end{eqnarray}

\section{Corner entropy at the second order}

In this appendix, we give some results for the logarithmic terms of EE at order $O(\lambda_m)^2$ for 3d CFTs. Solving eq.(\ref{firstintegral}) to the second order of $\lambda_m$ and then substituting the solution into the entropy functional eq.(\ref{HEEMiao3d3}), we can derive
\begin{eqnarray}\label{aHCGappendix}
a(h_0)
&=&(-\frac{f_0}{6}+4 c^2_1)a_E(h_0)+\sum_{m=2}^{\infty}\big(\lambda_m a_{K,m}(h_0)+\lambda_m^2 a_{K,(2) m}(h_0)\big)+O(\lambda_m)^3
\end{eqnarray}
where  $a_{(2)m}(h_0\to \infty)$ are given by
\begin{eqnarray}\label{amHCGappendix}
a_{K,(2)m}&=&-\frac{1}{2G}\int_{0}^{\infty}\frac{dy}{h_0^{4m-2}}\frac{2 m^2}{y^2 \left(y^2+1\right)^{4 m+1}}\bigg(512 m^4 \left(y^2+1\right)^2 y^4+123 y^8+312 y^6\nonumber\\
&&-64 m^3 \left(22 y^2+3\right) \left(y^3+y\right)^2-4 y^2 \left(\left(y^2+1\right)^{2 m}-12\right)+9 \left(\left(y^2+1\right)^{4 m}-1\right)+y^4 \left(246-4 \left(y^2+1\right)^{2 m}\right)\nonumber\\
&&-4 m \left(\left(y^2+1\right)^{2 m}+6 \left(y^2+1\right)^{4 m}-2 y^2 \left(\left(y^2+1\right)^{2 m}-28\right)-3 y^4 \left(\left(y^2+1\right)^{2 m}-102\right)+173 y^8+416 y^6-7\right)\nonumber\\
&&+16 m^2 \left(\left(y^2+1\right)^{4 m}-y^2 \left(\left(y^2+1\right)^{2 m}-24\right)-y^4 \left(\left(y^2+1\right)^{2 m}-143\right)+92 y^8+210 y^6-1\right)\bigg)\nonumber\\
&=&-\frac{1}{2G}\frac{\bar{a}_{(2) m}\pi}{h_0^{4m-2}}+O(\frac{1}{h_0^{4m}}).
\end{eqnarray}
 Here $\bar{a}_{(2) m}$ are some finite numbers:
\begin{eqnarray}\label{baramappendix}
\bar{a}_{(2) 1}=0,\ \bar{a}_{(2) 2}=-\frac{2825}{32},\ \bar{a}_{(2) 3}=-\frac{21845619 }{32768},\ \bar{a}_{(2) 4}=-\frac{657678125 }{262144},\ \bar{a}_{(2) 5}=-\frac{14523464909675 }{2147483648},...
\end{eqnarray}

Similarly, we can derive the opening angle $\Omega$ as
\begin{eqnarray}\label{angleHCGappendix}
\Omega=2\int_{h_0}^{0}dh\frac{1}{h'}=\Omega_E(h_0)+\sum_{m=2}^{\infty}\big(\bar{\lambda}_{m} \Omega_m(h_0)+\bar{\lambda}^2_{m} \Omega_{(2) m}(h_0)\big)+O(\bar{\lambda}_{m})^3
\end{eqnarray}
In the smooth limit $h_0\to \infty$, we have
\begin{eqnarray}\label{angleHCGmappendix}
\Omega_{(2) m}&=&\frac{2}{h_0^{4m-3}}\int_{0}^{\infty}dy\frac{4 m^2 (2 m-1) }{y^2  \left(y^2+1\right)^{4 m+1}}\bigg(4 (m-1) \left(y^2+1\right)^{4 m}+\left((3-4 m) y^2-1\right) \left(y^2+1\right)^{2 m+1}\nonumber\\
&&\ \ \ \ +\left(y^2+1\right)^2 \left((4 m-3) (16 m (2 m-3)+17) y^4-2 (8 m (3 m-5)+15) y^2-4 m+5\right)\bigg)\nonumber\\
&&+O(\frac{1}{h_0^{4m-1}})\nonumber\\
&=&  \frac{2\omega_{(2) m}\pi}{h_0^{4m-3}}+O(\frac{1}{h_0^{4m-1}})
\end{eqnarray}
with $\omega_{(2)m}$ given by
\begin{eqnarray}\label{omegamappendix}
\omega_{(2)1}=0,\ \omega_{(2)2}=-\frac{1359}{16},\ \omega_{(2)3}=-\frac{10742535 }{16384},\ \omega_{(2)4}=-\frac{325720423 }{131072},\ \omega_{(2)5}=-\frac{7216252470675 }{1073741824},...
\end{eqnarray}

Note that eqs.(\ref{amHCGappendix},\ref{angleHCGmappendix}) obey the power law eqs.(\ref{powerlawOmega},\ref{powerlawa}). Remarkably, the higher order terms $O(\lambda^n_m)$ behave as $O(1/h_0^{4mn})$ which are much smaller than the first order terms $O(\lambda_m)$ in the smooth limit $h_0\to \infty$. Thus it is sufficient to consider only the first order of $\lambda_m$ for the proof of the conjecture (\ref{conjecture}).

To end this section, let us comment the higher order terms $O(\lambda^n_m)$. 
At the first order $O(\lambda_m)$, all the possible terms of $(K^{2m})$ are equivalent to $ tr K^{2m}$ up to some factor. For the second order $\lambda^2_m$, $(K^{2m})$ can be classified by at most three equivalence classes, i.e., $tr K^{2m}$, $tr K^2 tr K^{2m-2}$ and $tr K^4 tr K^{2m-4}$. Similarly, more and more equivalence classes need to be considered for $(K^{2m})$ terms at higher order $O(\lambda^n_m)$. It is expected that all the equivalence classes obey the same power law  eqs.(\ref{powerlawOmega},\ref{powerlawa}) as $tr K^{2m}$. To see this, recall that there are two eigenvalues $k_{\pm}$ for the extrinsic curvature $K^i_j$ in $AdS_4$. Thus, we have 
\begin{eqnarray}\label{powerlawequivalenceclasses}
2^{n-1}tr K^{2m} &\ge& \prod_{i=1}^{n}Tr K^{2m_i}\ge tr K^{2m} ,\ \text{with} \sum_{i=1}^{n}m_i=m \nonumber\\
2^{n-1}(k^{2m}_++k^{2m}_-)&\ge&\prod_{i=1}^{n}(k^{2m_i}_++k^{2m_i}_-)\ge (k^{2m}_++k^{2m}_-).
\end{eqnarray}
Clearly, $\prod_{i=1}^{n}Tr K^{2m_i}$ are lower and higher bounded by $tr K^{2m}$ with some factors. So it is expected that $\prod_{i=1}^{n}Tr K^{2m_i}$ obey the same power law  eqs.(\ref{powerlawOmega},\ref{powerlawa}) as $tr K^{2m}$.

Let us take the $(K^4)$ terms of order $O(\lambda^2_2)$ as an example. These terms are associated to the cubic curvature gravity. There are two equivalence classes $tr K^{4}$ and $(tr K^2)^2$. We have studied $tr K^{4}$ above. For $(tr K^2)^2$, similar to eqs.(\ref{amHCGappendix},\ref{angleHCGmappendix}), we obtain
\begin{eqnarray}\label{amHCGappendix2}
a'_{K,(2) 2}
&=&\frac{1801}{16G}\frac{\pi}{h_0^{6}}+O(\frac{1}{h_0^{8}})\\
\Omega'_{(2) 2}
&=& -\frac{847\pi}{2h_0^{5}}+O(\frac{1}{h_0^{7}})
\end{eqnarray}
which indeed obey the same power law as $tr K^{4}$. One can further check that  $tr K^{4}$ and $(tr K^2)^2$ also obey the same power law  eqs.(\ref{powerlawOmega},\ref{powerlawa}) at the next order  $O(\lambda^3_2)$.

\end{document}